\documentclass[preprint2]{aastex}
\usepackage[usenames,dvips]{color}
\usepackage{graphicx,natbib}
\usepackage{hyperref}
\usepackage{amsmath}
\usepackage{multirow}

\newcommand{\altm} {\altaffilmark}

\shorttitle{X-ray pulsed detection of PSR J0007+7303}
\shortauthors{Lin et al.}

\begin{document}

\title{Detection of an X-ray Pulsation for the Gamma-ray Pulsar Centered in CTA 1}

\author{L. C.C. Lin\altm{1}, R. H. H. Huang\altm{2},  J. Takata\altm{3}, C. Y. Hwang\altm{1}, A. K. H. Kong\altm{2,5} and C. Y. Hui\altm{4}}

\email{lupin@crab0.astr.nthu.edu.tw}

\begin{abstract}

We report the detection of X-ray pulsations with a period of $\sim$315.87~ms from the 2009 {\it XMM-Newton} observation for the radio-quiet $\gamma$-ray pulsar, LAT PSR~J0007+7303, centered in the supernova remnant CTA~1.
The detected pulsed period is consistent with the $\gamma$-ray periodicity at the same epoch found with the $\it Fermi$ Gamma-ray Space Telescope.
The broader sinusoidal structure in the folded light curve of the X-ray emission is dissimilar to that of the $\gamma$-ray emission, and the phase of the peak is about $0.5$ shifting from the peak in the $\gamma$-ray bands, indicating that the main component of the X-rays originates from different sites of the pulsar.
We conclude that the main component of the X-ray pulsation is contributed by the thermal emission from the neutron star.
Although with a significantly different characteristic age, PSR~J0007+7303 is similar to Geminga in emission properties of X-rays and $\gamma$-rays; this makes PSR~J0007+7303 the second radio-quiet $\gamma$-ray pulsar with detected X-ray pulsations after Geminga.

\end{abstract}


\altaffiltext{1}{Graduate Institute of Astronomy, National Central University,
 Jhongli 32001, Taiwan}
\altaffiltext{2}{Institute of Astronomy and Department of Physics, National Tsing Hua University,
    Hsinchu 30013, Taiwan}
\altaffiltext{3}{University of Hong Kong,  Department of Physics, Hong Kong, PRC}
\altaffiltext{4}{Department of Astronomy and Space Science, Chungnam National University, Daejeon, South Korea}
\altaffiltext{5}{Golden Jade Fellow of Kenda Foundation, Taiwan}

\section{Introduction}

\begin{figure*}
\centering
\hspace*{\fill}\includegraphics[width=7.3cm]{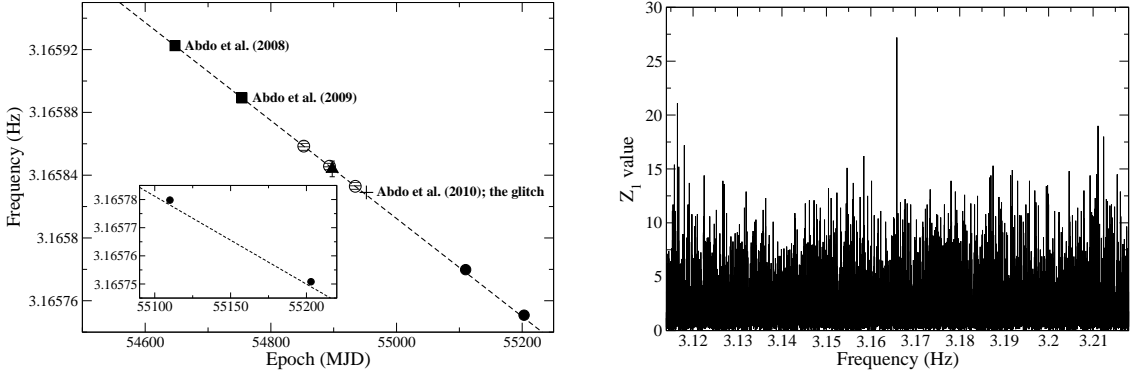}
\hspace*{\fill}\includegraphics[width=6.8cm]{PVLsearch.eps}
\hspace*{\fill} \caption{\small Left panel: Spin periods of
PSR~J0007+7303 from {\it Fermi} and {\it XMM}. The squares represent
the spin periods detected by \citet{Abdo2008,Abdo2009}.
 The plus
sign marks the glitch of PSR~J0007+7303 with $\rm{df/f}=5\times
10^{-7}$ \citep{Abdo2010}. The circles represent the spin
periodicities detected by the {\it Fermi} data before the glitch,
and the solid circles represent those after the glitch. The triangle
represents the period detected by the {\it XMM} data of 2009 with an
uncertainty shown by the error bar. The dashed line represents the
period derivative $-3.6133\times10^{12}$~Hz/s derived by
\citet{Abdo2009}. The periods after glitch have a little shift from
the dashed line which are clearly shown in the insert panel. Right
panel: $Rayleigh$ test of {\it XMM} data around the frequency
[3.114,3.218] Hz. The width of each independent trial is about
$1.037\times 10^{-5}$~Hz. The most significant signal appears at the
frequency of 3.165844 Hz.} \label{PDetection}
\end{figure*}

In the first year of its operation, {\it Fermi} observatory detected 23 new radio-quiet $\gamma$-ray pulsars \citep{Abdo2009,Saz2010}.
The X-ray counterparts for some of these $\gamma$-ray pulsars have been suggestively proposed, e.g., CXOU~J180950.2-233223  \citep{BR2002}, RX~J1836.2+5925 \citep{HGMC2002,HCG2007}, and 2XMM~J202131.0+402645 \citep{Tre2010}; these sources show similar spectral features as a neutron star.
Although these pulsars might be claimed as a next Geminga (e.g., \citealt{Abdo2010G}), their pulsed emission in X-ray bands have not been confirmed so far.

The X-rays from these radio-quiet $\gamma$-ray pulsars are expected to be generally characterized by thermal emission from the neutron star surface and the hot polar cap as well as non-thermal emission from the magnetosphere and the pulsar wind nebula \citep{Car2003,Car2004}.
These different emission components can possibly have their contributions in different rotational phases (cf. \citealt{Car2004}).
Therefore, it is not possible to obtain a complete picture of this unique class of pulsars without the detection of X-ray pulsations.
Nevertheless, until now, Geminga is still the only one in this class with the X-ray pulsed signals identified.

In the search for X-ray pulsations from $\gamma$-ray-only-emitting pulsars, the point source centered in the composite supernova remnant (SNR) CTA~1 (G119.5+10.2) is one of the most intriguing candidates.
CTA~1 was firstly discovered by {\it Owens Valley Radio Observatory} in 1960s \citep{HR60}; previous studies \citep{SSS95} revealed that the X-rays from the center of CTA~1 consist of diffuse emission plus a faint point source RX~J0007.0+7302.
The precise position of RX~J0007.0+7302 was provided by the {\it Chandra} image \citep{HGCH2004}.
No radio counterpart has been found at the position of RX~J0007.0+7302 \citep{Pin93}.
On the other hand, the high $\gamma$-ray-to-X-ray and X-ray-to-optical flux ratios \citep{Bra98,HGCH2004} and the morphology of a bent jet from a torus-like pulsar wind nebula (PWN) revealed in X-ray image \citep{HGCH2004} all characterize RX~J0007.0+7302 as a neutron star.
Further studies of the spectral properties for RX~J0007.0+7302 were also consistent with a pulsar without radio emission, e.g., Geminga \citep{Sla2004}.
In addition, based on the positional coincidence and the emission properties, RX~J0007.0+7302 was also identified as the X-ray counterpart of 3EG~J0010+7309, which is a $\gamma$-ray source found by the Energetic Gamma-Ray Experiment Telescope (EGRET) on the {\it Compton Gamma-Ray Observatory} \citep{Sla97,Bra98}.

Although all these observations suggested that RX~J0007.0+7302 is likely to be a radio-quiet $\gamma$-ray pulsar associated with the supernova remnant \citep{Sla97}, the final confirmation came from the detection of $\gamma$ pulsations with {\it Fermi} \citep{Abdo2008}.
The detected $\gamma-$ray pulsation has a spin period of $315.86$~ms.
This pulsar has a characteristic spin-down age of $\tau\sim 14000$~yr, which is consistent with the age of CTA~1 (10000 -- 20000~yr, \citealt{ZC98} and \citealt{Sla2004}).
This further supports the scenario that this pulsar is indeed associated with CTA~1.
In addition, a precise position of LAT~PSR~J0007+7303 can be inferred from the {\it Chandra} image \citep{HGCH2004,Abdo2009}.
However, the search of the corresponding pulsed signal in the archival X-ray data obtained before 2009 was unsuccessful.
\citet{Car2009} indicated that the X-ray pulsation can be derived from the contemporaneous {\it Fermi} ephemeris of PSR~J0007+7303 although no any further information of the X-ray pulsation was provided.
This leaves the X-ray temporal properties of RX~J0007.0+7303 still unconstrained.

In this Letter, we report the detection of X-ray pulsations from
PSR~J0007+7303 using the 2009 {\it XMM-Newton} (hereafter {\it XMM})
observation (PI: \citealt{Car2008}, ObsID: 06049401). We show
detailed information of temporal and spectral analyses for this
source. We investigate the source spectrum and the folded light
curves in different energy bands to determine the possible origins
of the X-ray emission from PSR~J0007+7303.

\section{Observations and data analysis}

The central region of CTA~1 was observed by {\it XMM} twice with an epoch separation of $\sim$ 7 years.
The first observation was performed on 2002 February 21 for $\sim$ 40 ks and the second one was carried out on 2009 March 7 for $\sim$ 120 ks.
These two observations were operated in the full-frame mode for the two MOS detectors and in the small-window mode for the pn detector.
The medium and thin filters were used for the two MOS detectors and the pn detector, respectively.
The temporal resolution of $\sim$6~ms for the small-window mode of pn detector is suitable for the periodicity search of X-ray pulses from PSR~J0007+7303.
\citet{Sla2004} have already analyzed the old {\it XMM} data observed on 2002 February 21 but no significant pulsation was detected using a total of 1055 counts in the energy band 0.3--10 keV and an upper limit  of 61\% for the pulsed fraction was reported with a sinusoidal profile.
In our study, we focus on the 2009 XMM data.
All the data were processed with the SAS version of 10.0.0.

\subsection{Timing analysis}

We first performed the source detection by using the task  ``edetect$\_$chain'' and determined the position of PSR~J0007+7303 in the X-ray band at (J2000) R.A.=$00^h07^m01^s.2$, Dec.=$+73^{\circ}03'07''.6$ with an uncertainty of $\sim 0''.6$.
The point source was then extracted at this position within a circular region of 15$''$ in radius, which is consistent with the 70\% of the encircled energy function.
After screening the background flare, the effective exposure is $\sim$ 67 ks.
Following the standard process of the data reduction, 2989 counts were yielded in the energy band 0.2--12 keV.
Furthermore, the photon arrival times were corrected to the solar system barycenter with the ``barycen'' task (JPL DE200 earth ephemeris) of the SAS package.

\begin{figure}[t]
\centering
\includegraphics[width=6.2cm]{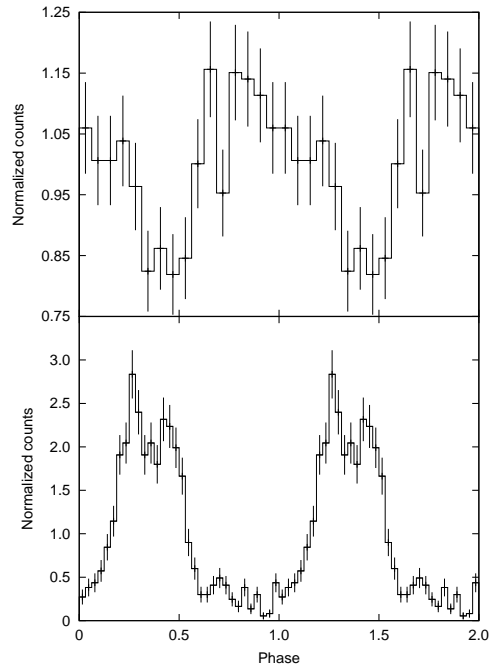}
\caption{{\small Folded light curves of PSR~J0007+7303. The upper panel shows the pulse profile in the 0.2-12 keV from the 2009 {\it XMM} observation and the lower one shows the profile in the 100 MeV to 300 GeV from the {\it Fermi} observations.
The counts were normalized to the average photons of each bin.
All the curves were folded simultaneously with the periodic frequency and the frequency derivative of the same epoch at the start of GTI of the 2009 {\it XMM} data as we described in the main text.}}

\label{PFGX}
\end{figure}

\citet{Abdo2009} reported a periodic signal of 3.1658891845(5) Hz at the reference epoch MJD 54754 with a frequency derivative of $-3.6133(3)\times 10^{-12}$~Hz/s for PSR~J0007+7303. 
According to this timing ephemeris, a periodicity of 3.165844344(4) Hz is expected at the start of the GTI of the 2009 {\it XMM} data (epoch MJD 54897.6335129). 
We thus search for periodic signals with the known frequency derivative around the predicted frequency in the X-ray data.
By performing an $H$-test \citep{DSR89}, we found a test statistic of $H\sim 27.0$, which corresponds to a random probability
of $2.0\times 10^{-5}$ for a single trial \citep{DB2010}. 
The uncertainty of our detection was determined by the equation provided by \citet{Lea87} and the most significant trial for the periodicity search was determined at 3.165844(5) Hz as shown in the left panel of Fig.~\ref{PDetection}. 
The $H$-statistic also obtains the most significant periodic detection with a sinusoidal structure consisting of only one harmonic. 
We then verified the $Z_1$ value of ~10000 independent nearby trials ($\sim$ 3.114 Hz -- 3.218 Hz) by a $Rayleigh$ test. 
Our detection of the periodic signal is highly significant as shown in the right panel of Fig.~\ref{PDetection}.

In order to compare the pulse profiles in X-rays and in $\gamma$-rays, we obtain $\sim$ 3-month {\it Fermi} data (2009 January 21 -- 2009 April 13) within the energy range of 100 MeV to 300 GeV on the epochs close to the 2009 XMM observation. 
We selected a 1-degree radius region centered on the source at R.A.=$1.7565^{\circ}$, Dec.=$73.0523^{\circ}$ (J2000) according to
the position of the X-ray counterpart detected by {\it Chandra}. 
The pulse profiles folded simultaneously with the frequency of 3.165844344 Hz and the frequency derivative of $-3.6133(3)\times
10^{-12}$~Hz/s at epoch MJD 54897.6335129 for the {\it XMM} and the {\it Fermi} observations are shown in Fig.~\ref{PFGX}. 
The pulse profile in the X-ray band shows a large pulsed phase (0.5625 -- 1.3125) with 0.75 duty cycle while in the $\gamma$-ray band the pulse profile shows a small pulsed phase (0.125 -- 0.5625) with $\sim 0.45$ duty cycle. 
In addition, the peaks of these two light curves shift about 0.5~phase, indicating that the X-rays and $\gamma$-rays originate from different sites of the pulsar.

\subsection{Spectral analysis}

The EPIC spectra of PSR~J0007+7303 were extracted using the aforementioned 15$''$-radius source region.
The background was defined as an annulus with the inner and outer radii of 75$''$ and 120$''$ centered at the source region.
The photon redistribution matrices and the ancillary region files were generated by the task ``rmfgen'' and ``arfgen'' of the SAS package.
In order to compare with the previous analysis reported by \citet{Sla2004,HGCH2004}, our spectra were restricted in the energy range $0.5-10.0$~keV and each spectral bins were rebinned with a minimum of 30 counts.
To reduce the uncertainties in spectral fitting, we set the column density to be $\sim 2.8\times 10^{21}$~cm$^{-2}$ according to the previous {\it ASCA}/{\it ROSAT} measurements of CTA~1 \citep{Sla97}.
This value is also consistent with the optical extinction \citep{HGCH2004}.

\begin{table}[t]
\caption{\footnotesize{Best-fit spectral parameters for PSR J0007+7303.}}\label{spectrum}
\resizebox{1.05\columnwidth}{!} {
\begin{tabular}{l|cc}
\hline
Parameter & PL+BBodyrad & PL+NSA$^{c}$
\\
\hline
$N_H$ (cm$^{-2}$)  & $2.8 \times 10^{21}$ (fixed) & $2.8 \times 10^{21}$ (fixed)
\\
$\Gamma$ & 1.52$^{+0.10}_{-0.09}$  & 1.49$^{+0.10}_{-0.11}$
\\
$F_{X,\rm{PL}}$~(ergs cm$^{-2}$ s$^{-1}$)$^{a}$ & $1.6\times 10^{-13}$ & $1.6\times 10^{-13}$
\\
$kT$ (keV) & 0.104$\pm$0.013 & 0.058$^{+0.019}_{-0.011}$
\\
$F_{X,\rm{Th}}$~(ergs cm$^{-2}$ s$^{-1}$)$^{a}$ & $3.3\times 10^{-14}$ & $3.5\times 10^{-14}$
\\
$R$~(km)$^{b}$ & 1.39$^{+0.68}_{-0.43}$ & 10 (fixed)
\\
$\chi^2$/dof & 108.9/119 & 108.0/119
\\
\hline
\end{tabular}
}
\begin{minipage}[t]{0.98\linewidth}
\footnotesize{
Note: Quoted errors indicate the 90\% confidence level for one parameter of interest. \\
\noindent
${}^{a}$ The unabsorbed flux is measured in the range of 0.5--10 keV. \\
${}^{b}$ The radius is measured from the normalization factor for a source distance of 1.4 kpc.\\ 
${}^{c}$ The NSA model is proposed to be nonmagnetic with the mass and radius of the neutron star fixed at $1.4~\rm{M_{\odot}}$ and 10~km.
}
\end{minipage}
\end{table}

We have examined the spectrum by fitting with various single component models that include power-law (PL), blackbody(BB) and hydrogen atmospheric model of a neutron star (NSA; \citealt{ZPS96}).
We find that no single component can fit the spectrum well.
Instead, the spectrum can be described by a composite model consisting of a non-thermal component and a thermal component.
We summarize the best-fit spectral parameters in Table~\ref{spectrum}.
The non-thermal X-ray emission from the pulsar PSR~J0007+7303 is thought to come from the outer magnetospheric gap \citep{TSHC2006,TCS2008}.
The thermal contribution with a temperature of $\sim 0.6-1.0 \times 10^{6}$~K might originate from the surface of the neutron star.
The X-ray flux from the thermal component is only $\sim$20\% of the total flux, which is consistent with previous {\it XMM} detection \citep{Sla2004}.
However, the observed {\it XMM} flux is $\sim$ 4 times of the flux attributed to the pulsar in the {\it Chandra} observation \citep{HGCH2004}.
We note that the aperture radius of {\it XMM} is about 15$"$, which is much larger than the point-spread-function of {\it Chandra} and might contain emission from the PWN or/and the jet.
The excess of the thermal flux detected by {\it XMM} data may be also due to the thermal emission contributed from the diffuse emission of CTA~1 \citep{Sla97}.
We also note that the {\it Chandra} spectrum was fitted with only 6 dofs, suggesting the model fitting might have large uncertainty.
In this respect, the uncertainty of the thermal flux derived by the {\it Chandra} spectrum might be large.
On the other hand, although the resolving power of {\it XMM-Newton} is worse, the larger effective area of the detectors onboard {\it XMM-Newton} and the longer exposures of the 2009 {\it XMM} data should provide much better photon statistics than that of the {\it Chandra} data.
Furthermore, we note that the X-ray light curve shows that the pulsed components can be as large as $\sim$ 20\% of the DC level, indicating that the soft X-ray flux from the pulsar could be at least as large as $\sim$ 20\% of the detected flux even the observed total flux might be contaminated by PWN or jets around the pulsar.
We also note that the sum of the individual non-thermal contributions from the jet, PWN and the pulsar reported by \citet{HGCH2004} is fully consistent with that inferred from the {\it XMM} data, though it is not possible to disentangle various components with this data set.
These indicate that the fluxes of these two data sets are not inconsistent.

\begin{figure}[t]
\centering
\caption{{\small Broad band pulsed spectrum of PSR~J0007+7303.}}
\includegraphics[width=7.5cm]{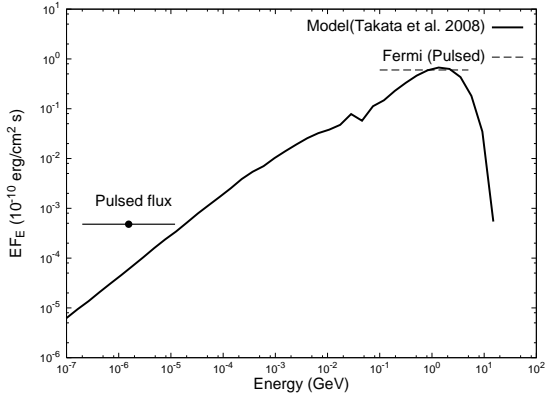}
\label{spec}
\end{figure}


\section{Discussion}

An X-ray pulsation was detected using 2009 {\it XMM-Newton} data.
However, it is difficult to obtain a pulsed spectrum with present data.
The photons of the pulsed component are quite few with the pulsed fraction $\sim$ (17$\pm$1)\% of the total X-ray flux.
In addition, the EPIC background is dominated by thermal emission at lower energies (E~$<$~1~keV), which causes large uncertainties in the pulsed spectrum.
For $\gamma$-ray pulsars in general, however, the X-ray emission can be fitted by a thermal component from the neutron star surface and/or a non-thermal component from the relativistic particles that also emit the $\gamma$-rays.
It has been observed from several $\gamma$-ray pulsars (e.g., Geminga and Vela) that the peak in X-ray light curve of the non-thermal component is  in phase with the $\gamma$-ray peak, while the peak phase of the thermal component  is different from that in the $\gamma$-ray bands.
The $\sim$0.5 phase shift between the peaks of the folded light curves in the X-ray and the $\gamma$-ray bands, as shown in Fig.~\ref{PFGX}, indicates that the main pulsed component of the X-rays probably originates from the thermal emission of the neutron star.

\begin{figure}[t]
\centering
\includegraphics[width=6.2cm]{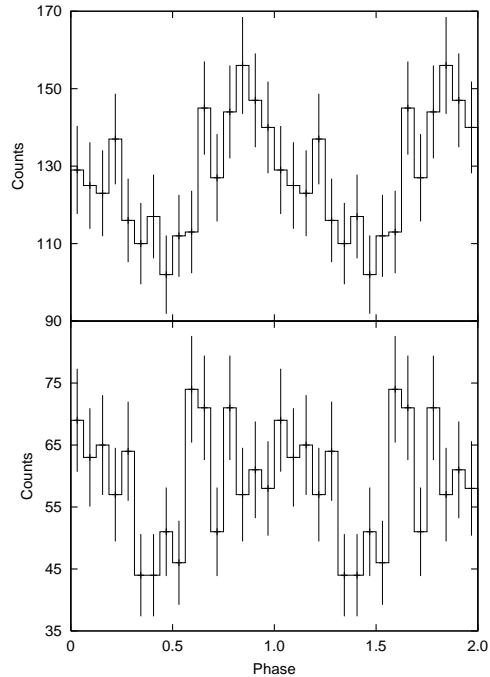}
\caption{{\small Folded light curves of PSR~J0007+7303 in the 0.2--2~keV (upper panel) and in the 2--12~keV (lower panel). These two curves were folded with the frequency of 3.165844344~Hz and the frequency derivative of $-3.6133\times 10^{-12}$~Hz/s
simultaneously at the same epoch zero of MJD 54897.6335129.}}
\label{PFDX}
\end{figure}


In Fig.~\ref{spec}, we compare the observed pulsed spectrum with the theoretical predictions of the outer gap accelerator model investigated by \citet{TCS2008}, where the $\gamma$-rays and the X-rays are emitted via curvature radiation and synchrotron radiation of the relativistic particles in the magnetosphere, respectively.
We can see from Fig.~\ref{spec} that the observed pulsed flux integrated from 0.2 to 12~keV of $\sim (5\pm 0.2) \times10^{-5}~\mathrm{photons~cm^{-2}~s^{-1}}$ is much larger than the predicted pulsed flux ($\sim 10^{-5}~\mathrm{photons~cm^{-2}~s^{-1}}$) of the non-thermal component, suggesting the observed pulsed component is not dominated by the non-thermal emission from the outer gap accelerator.

In order to examine the energy dependence of the pulsed X-ray emission from PSR~J0007+7303, we divided the source photons into the soft X-ray (0.2 -- 2 keV) and hard X-ray (2 -- 12 keV) band.
Fig.~\ref{PFDX} shows the folded light curves in the soft (upper panel) and in the hard (lower panel) X-ray bands.
For the soft X-ray energy band, the random probability to yield the pulsed detection is $3.1\times 10^{-5}$ ($Z_1^2=20.8$), and
the pulse profile has a broad sinusoidal structure, which shows a similar X-ray pulse profile as that in Fig.~\ref{PFGX}.
In the hard X-ray band, on the other hand, the random probability to yield the pulsed detection is only 0.039 ($Z_1^2=6.49$), and the structure of the pulse profile is much more random.
We did not detect pulsed signals at hard X-rays.
This suggests that the X-ray pulsation detected for PSR~J0007+7303 is mainly from the thermal emission.
This is also consistent with the fact that the thermal X-ray emission from a neutron star, which may not contribute significantly to the hard X-rays ($>2~$keV).
Without a firm detection of pulsed emission in the hard X-ray band, the nature of the power-law component remains unclear.

One can use some physical properties of pulsars such as ages and spin-down powers to characterize features of $\gamma$-ray pulsars.
Based on these physical properties, PSR~J0007+7303 would be considered to be similar to a Vela-like pulsar.
On the other hand, one can also consider the similarity between two pulsars by comparing their high-energy emission properties.
In this respect, PSR~J0007+7303 has the certain X-ray pulsed detection and shares the similar spectral features with the Geminga pulsar; this indicates that the high-energy emission from these two pulsars may have similar mechanisms and makes PSR~J0007+7303 as the second Geminga.
Both the spectra of PSR~J0007+7303 and Geminga show a steepening at the range of GeV and can be described as a hard power-law with $\Gamma\sim 1.5$ -- 1.6 \citep{May94,Abdo2008}.
The X-ray source spectra of these two pulsars can both be fitted by a composite model consisting of a power-law and a blackbody component \citep{HW97,Sla2004}.

However, in contrast to the Geminga pulsar, the X-ray spectral behavior of PSR~J0007+7303 is dominated by non-thermal contribution, which might come from the PWN \citep{Sla97}.
Since the pulse profile of the pulsar centered in CTA~1 presents a broad sinusoidal structure as that of the Geminga pulsar \citep{Car2004}, the pulsation in the soft X-ray band is expected to be thermal-dominant.
Both sources show significantly dissimilar pulsed profiles for the soft X-ray and the $\gamma$-ray bands (\citealt{HR93} \& this work).
In fact, except for the Crab pulsar, most of the {\it EGRET} pulsars do not show resemblance for the pulse profiles in the soft X-ray and the $\gamma$-ray bands \citep{Thp2001}.
The hard X-ray and the $\gamma$-ray pulsed profiles of the Geminga pulsar have similar features \citep{KPZR2005}, indicating the emission are likely originated from the same locations.

\acknowledgments

The authors appreciate an anonymous referee for his/her fruitful comments.
We also thank Dr. Jau-Shian Liang for discussion on the application of the pulsed spectrum for our target.
This work was partially supported by the National Science Council through grant NSC 99-2811-M-008-057, and
RHHH is supported through grant NSC 099-2811-M-007-062.
CY Hwang acknowledges support from the National Science Council through grants
NSC~99-2112-M-008-014-MY3 and NSC~99-2119-M-008-017.
CY Hui is supported by the research fund of Chungnam National University in 2010.
AKHK acknowledges support from the National Science Council through grants NSC~96-2112-M-007-037-MY3 and NSC~99-2112-M-007-004-MY3.

{\it Facilities:} \facility{Fermi}, \facility{XMM-Newton}.

\end{document}